\RequirePackage{fix-cm}
\documentclass[smallextended]{svjour3}       % onecolumn (second format)
\smartqed  % flush right qed marks, e.g. at end of proof
\usepackage{graphicx}
 \usepackage{mathptmx}      % use Times fonts if available on your TeX system
\journalname{J Low Temp Phys}

\begin{document}
\title{Use of two-body correlated basis functions with van der Waals interaction to study the shape-independent approximation for a large number of trapped interacting bosons.}

\titlerunning{On the large number of trapped interacting bosons.}        % if too long for running head

\author{M.L. Lekala \and B. Chakrabarti \and T.K. Das \and G.J. Rampho \and S.A. Sofianos \and R.M. Adam \and S. K. Haldar}

\institute{M.L. Lekala \and B. Chakrabarti \and G.J. Rampho \and S.A. Sofianos\thanks{deceased} \at Department of Physics, University of South Africa, P.O.Box 392, Pretoria 0003, South Africa. 
\and 
B. Chakrabarti \at \emph{Permanent address: Department of Physics, Presidency University, 86/1 College Street, Kolkata 700 073, India.}
\and
G.J. Rampho \at Tel.: +27 12 429 8640\\ \email{ramphogj@gmail.com}   
\and
T.K. Das \at Department of Physics, University of Calcutta, 92 A.P.C. Road, Kolkata 700084, India
\and
R.M. Adam \at Square Kilometer Array Radiotelescope, The Park, Park Rd., Pinelands 7405, South Africa
\and 
S. K. Haldar \at Theoretical Physics Division, Physical Research Laboratory, Navrangpura, Ahmendabad-380 009, India, \at {Department of Physics, University of Haifa, Haifa, Israel. }}

\authorrunning{Lekala et al} % if too long for running head

\date{Received: date / Accepted: date}

\maketitle
\begin{abstract}
  We study the ground state and the low-lying excitations of a trapped Bose gas in an isotropic harmonic potential for very small ($\sim 3$) to very large ($\sim 10^7$) particle numbers. We use the correlated two-body basis functions and the shape-dependent van der Waals interaction in our many-body calculations. We present an exhaustive study of the effect of inter-atomic correlations and the accuracy of the mean-field equations considering a wide range of particle numbers. We calculate the ground state energy and the one-body density for different values of the van der Waals parameter $C_{6}$. We compare our results with those of the modified Gross-Pitaevskii results, the correlated Hartree hypernetted-chain equations (which also utilize the two-body correlated basis functions), as well as of the Diffusion Monte Carlo for hard sphere interactions. We observe the effect of the attractive tail of the van der Waals potential in the calculations of the one-body density over the truly repulsive zero-range potential as used in the Gross-Pitaevskii equation and discuss the finite-size effects. We also present the low-lying collective excitations which are well described by a hydrodynamic model in the large particle limit.
\end{abstract}

\section{Introduction}

Laboratory realization of gaseous Bose-Einstein condensates (BEC) \cite{Anderson,Bradley,Davis} and subsequent experiments characterizing low-lying collective excitations~\cite{Jin,Jin1} have prompted various theoretical investigations of these systems \cite{Stringari,Dalfovo1,Hu,Andras,Fil,JP}. These theoretical calculations include the trapped Fermi gas and BEC in a shallow trap. Unlike phenomena such as superfluidity of liquid helium, the atomic vapor is very dilute where the fundamental interactions are characterized by the $s$-wave scattering length $a_s$. With this type of interactions the system is quite easy for theoretical understanding. The standard theory for BEC in a dilute atomic vapor uses the Gross-Pitaevskii (GP) equation~\cite{Dalfovo} which is a nonlinear Schr\"odinger equation with the inter-atomic interaction characterized by the $s$-wave scattering length $a_s$. The exact shape of the inter-particle interaction and inter-atomic correlation are ignored in this picture. The elementary excitations of a BEC in a harmonically confined dilute Bose gas have been studied using the GP equation~\cite{Dalfovo2}.

The GP theory is the most popular tool for the description of weakly interacting bosons. However, in recent experiments the number $N$ of trapped atoms varies from just a few to $\sim 10^7$. Therefore, it has become imperative to study the effect of realistic interaction, inter-atomic correlation and the accuracy of the GP equation. The astonishing success of the mean-field theory lies in the fact that the gas parameter $n\,a_{s}^{3}$ ($n$ is the atom number density) is small. However, nowadays, utilizing the Feshbach resonances, $a_{s}$ can be tuned by changing the magnetic field. Particle-particle correlation becomes important at large $a_s$ and the accuracy of the GP theory needs a deeper study. The mean-field description uses an effective mean-field potential obtained by the shape-independent pseudopotential approximation (SIA). The SIA implies that the calculated ground state energy remains unchanged irrespective of the shape of the interaction potential. As the hard sphere interaction is completely characterized by a single parameter, it is beyond the capacity of characterizing the universal behavior of the ground state properties. Instead, we utilize a realistic interaction having a controllable parameter $(C_{6})$.

The SIA has been addressed in different context. For homogeneous systems, Cowell {\it et al}~\cite{Cowell} have shown that the SIA fails when $n\,a_{s}^{3} > 0.5 $, i.e., different potentials having the same scattering length leads to different ground state energies. In the case of an inhomogeneous system with few atoms in the trap, the SIA is less valid for tight confinement. For a homogeneous Bose gas, Giorgini {\it et al.}~\cite{Giorgini} find a small dependence of the ground state energy on the exact shape of the two-body potential when $n\,a_{s}^{3} \simeq 10^{-3}$. For larger $n\,a_{s}^{3}$ or for stronger confinement the SIA becomes less applicable. The validity of the SIA has also been addressed for both weakly and strongly interacting BECs. The ground state energy of a trapped BEC is calculated by the diffusion Monte Carlo (DMC) approach using different two-body potentials that generate identical $a_{s}$~\cite{Blume}. It is seen that different potentials produce indistinguishable total ground state energies for a small gas parameter. Whereas for larger $a_{s}$, inter-atomic correlations play an important role and SIA becomes invalid and quantum corrections to the ground state energy are required~\cite{Blume}. The local density versus correlated basis approaches for trapped bosons have been examined and the beyond the GP approximation has been prescribed~\cite{Poll,Poll1}

In the present manuscript, as stated before, we report the ground state properties of weakly interacting systems with a wide range of $N$ using the two-body correlated basis functions. Although the ground state properties have been addressed in different theories as described above, we do not find any exhaustive study which keeps the effect of inter-atomic correlations, uses the realistic van der Waals interaction and treats the real experimental situation where the number of particles is finite, varying from very few to a quite large number. Thus, the first key question of our study is to test the accuracy of the mean-field theory for large $N$ and to study the SIA for a wide range of $N$. As the modified GP (MGP) equation accounts for quantum fluctuations, we also check how closely our correlated basis functions reproduce the ground state energy obtained with the MGP. We also compare our calculated results with the correlated Hartree HNC results~\cite{Poll} which utilizes hard sphere bosons. We observe that the MGP substantially improves the GP results and our correlated many-body energies are in good agreement with the MGP. All the earlier calculations in this direction consider a truly finite number of particles and use the standard short-ranged two-body potential. Here we use the realistic van der Waals potential with a short-ranged hard core and a long-ranged attractive tail characterized by the parameter $C_{6}$. Comparison with the DMC calculation is also made for few bosons with a repulsive hard sphere interaction to justify the accuracy of our two-body correlated basis function in the dilute regime since the DMC is essentially exact. Thus, we are further assured to tackle real experimental situations which consider quite a large number of bosons and, hence, is beyond the scope of the DMC.

We investigate the effect of the long attractive tail (by changing $C_{6}$, adjusting the cutoff radius $r_{c}$, such that they produce the same scattering length) on the ground state energy for a wide range of $N$. For $a_{s} = 100\,{\rm Bohr}$, we observe good agreement between our many-body results and the mean-field results. This establishes the applicability of the SIA for a wide range of $N$. In addition to the ground state energy, we also calculate the effect of the long-range attractive tail of the van der Waals interaction on the one-body density for a small and a large number of atoms in the trap. Although we observe that the one-body density profile is almost independent of the choice of the long-range attractive tail of the van der Waals potential, the peak value of the density substantially differ from the mean-field results. We next analyze the finite size effect on the one-body density profile.

Studies of collective excitations at both low and high energies at the large-$N$ limit is quite interesting for the following reason. The excitations at high energies are expected to be of single-particle nature. However, a laboratory BEC is a strongly inhomogeneous system due to the external trap and may significantly differ from the uniform Bose gas at low energies where only phonons are present. Thus the study of the transition from low-lying collective excitations to single-particle excitations by the correlated many-body method using the realistic interaction is itself interesting. In the present many-body calculation, we can in principle calculate all the low-lying and high-lying collective excitations. However, as mentioned in the formalism in the next section and in the result section, we solve the coupled differential equations by the hyperspherical adiabatic approximation (HAA) and consider the lowest eigen potential as the many-body effective potential. HAA with the lowest eigen potential is very accurate for the ground and few low-lying collective excitations. However, for the high-lying excitations the effect of higher order eigen potential will come in the picture and one should use the coupled adiabatic approximation. Thus for our present calculation we report only on the few low-lying excitations where the effect of higher order eigen potentials can be safely ignored. 

The paper is organized as follows. In Sec.~\ref{sec:method} we discuss the correlated potential harmonic basis for a large particle number. Sec.~\ref{sec:egs-largeN} mainly considers ground-state properties of the systems and comparison with the mean-field results is provided. The validity of the shape-independent approximation is presented in Sec~\ref{sec:SIA}. Sec.~\ref{sec:collective} deals with the calculation of collective excitations at low energies. Sec.~\ref{sec:conclusion} concludes with a summary.

\section{Formalism}
\label{sec:method}
\subsection{The correlated potential harmonic expansion method}

In the present work we calculate the ground state energy and the low-lying collective excitation frequencies of a dilute BEC for a large number of bosons ($\sim 10^{7}$) using the potential harmonics expansion method (PHEM)~\cite{Fabre83}. The most fundamental feature of the PHEM is that the two-body correlations are dominant in the many-body system and the two-body Faddeev component of the $N$-body wave function is a function of the two-particle relative separation $\vec{r}_{ij}$ and a global length called hyperradius $r$. We have already successfully utilized the two-body correlated basis functions for the description of dilute BECs~\cite{TKD}. So in the present description we only point out the essential components of the PHEM for the completeness and clarity of the manuscript. For a detailed formulation we refer the reader to Refs.~\cite{Das1,Das2,PLA2009}. 

In the PHEM we expand the two-body Faddeev component $\phi_{ij}$ corresponding to the $(ij)$ interacting pair of bosons, in a condensate with $N={\mathcal{N}} +1$ bosons, in terms of the potential harmonic (PH) basis as
\begin{equation}
\phi_{ij}(\vec{r}_{ij},r)
=r^{-(\frac{3{\mathcal N}-1}{2})}\sum_{K}\,{\mathcal P}_{2K+\ell}^{\ell m_\ell}
     (\Omega_{\mathcal N}^{ij})\,u_{K}^{\ell}(r)\, ,
\label{eq.faddeev-expansion}
\end{equation}
where $\Omega_{\mathcal N}^{ij}$ corresponds to the full set of hyperangles in the $ij$-th partition while $\ell$ and $m_\ell$ are the orbital angular momentum of the system and its projection. In the hyperspherical coordinate, the variables are characterized by the hyperradius $r^{2} = \sum_{i=1}^{\mathcal N}\zeta_{i}^{2}$ ($\vec{\zeta}_i$, $i=1,{\mathcal N}$ being the Jacobi vectors describing the relative motion) and $(3{\mathcal N}-1)$ hyperangles ~\cite{Ballot}. However, for the potential harmonic expansion method of a weakly interacting BEC, as the only the two-body correlations are dominating, we assume that when the $(ij)$ pair of atoms interact, the rest of the atoms are inert spectators~\cite{Das1}. As all the degrees of freedom coming from the $({\mathcal{N}}-1)$ inert spectators are frozen, the number of quantum numbers becomes effectively four irrespective of the number of bosons. These are the orbital angular momentum $\ell$, the azimuthal $m_\ell$, the grand orbital angular momentum $2K+\ell$, and the energy quantum number. The closed analytic expression for the PH, ${\mathcal P}_{2K+\ell}^{\ell m_\ell}(\Omega_{\mathcal N}^{ij})$ can be found in Ref.~\cite{Fabre83}. It is indicated in our earlier works that the above expansion is in general very slow as the lowest order PH is a constant and does not represent the strong short-range repulsion of the inter-atomic interaction. Therefore, we introduced a short-range correlation function $\eta(r_{ij})$ which is obtained as the zero-energy solution of the two-body Schr\"odinger equation 
\begin{equation}
 \Big[-\frac{\hbar^2}{m}\,\frac{1}{r_{ij}^2}\,\frac{{\rm d}}{{{\rm d}}r_{ij}}\left(r_{ij}^2\frac{{\rm d}}{{{\rm d}}r_{ij}}\right)+V(r_{ij})\Big]\eta(r_{ij})=0\, ,
\hspace*{.1cm}\,
\label{eq.tbse}
\end{equation}
with the chosen two-body potential $V(r_{ij})$~\cite{PLA2009}, and corresponds to the appropriate $s$-wave scattering length $a_s$ as described in next section. After including the correlation function in the PHEM basis, we call it the correlated potential harmonic expansion method (CPHEM). The expansion Eq.~(\ref{eq.faddeev-expansion}) now takes the form
\begin{equation}
\phi_{ij}(\vec{r}_{ij},r)=r^{-(\frac{3{\mathcal N}-1}{2})}\,\sum_{K}\,{\mathcal P}_{2K+\ell}^{\ell m_\ell}(\Omega_{\mathcal N}^{ij})\,u_{K}^{\ell}(r)\eta(r_{ij})\,. \label{eq.faddeev-corr-exp}
\end{equation}
Substituting this expansion (Eq.~(\ref{eq.faddeev-corr-exp})) in the many-body Schr\"odinger equation, one gets a set of coupled differential equations (CDE) in $r$. The coupling potential matrix element $V_{KK^{\prime}}$ is given by
\begin{eqnarray}
   V_{KK^{\prime}}(r) = & \frac{1}{\sqrt{h_{K}^{\alpha,\beta}\,
    h_{K^{\prime}}^{\alpha,\beta}}} \,\int_{-1}^{+1} \, P_{K}^{\alpha, \beta}(z)
    \,V\left(r\sqrt{(1+z)/2}\right) \nonumber\\ 
 & \times\,P_{K^{\prime}}^{\alpha, \beta}(z) \,\eta\left(r\sqrt{(1+z)/2}\right)\,W_{\ell}(z)\, {\rm d} z\,,
\label{corrPME}
\end{eqnarray}
where $P_{K}^{\alpha, \beta}(z)$, $h_{K}^{\alpha,\beta}$ and $W_{\ell}(z)$ are the Jacobi polynomial, its norm and weight function respectively, with $\alpha=(3N-8)/2$ and $\beta=\ell+\frac{1}{2}$. The CDEs are solved using the hyperspherical adiabatic approximation (HAA)~\cite{HAA}. The HAA basically reduces the whole $3{\mathcal{N}}$ dimensional problem to an effective one-dimensional problem. In the HAA, the coupled potential matrix $V_{KK^{\prime}}$ along with the diagonal hypercentrifugal repulsion is diagonalized to get an effective potential $\omega_0(r)$ as the lowest eigen value of the matrix for a particular value of $r$.

The basic length scale for a harmonic oscillator trap of frequency $\omega_{ho}$ is $a_{ho} = \sqrt{\hbar/(m\,\omega_{ho})}$. For the typical experimental BEC, $a_{ho}$ is of the order of $10^{4}\,a_0$. However, the effective potential in hyperspherical space due to the hypercentrifugal repulsion together with the harmonic oscillator trap has a minimum at about $\sqrt{3N}\,a_{ho}$. As an example, with $N \simeq 10^{4}$, the minimum of the effective potential will be near $10^{6}\,a_0$, which is almost $10^{5}$ times larger than the typical range of the inter-atomic interaction. This shows that for such a typical case, the entire contribution to $V_{KK^{\prime}}(r)$ in the integral in Eq.~(\ref{corrPME}) comes from an extremely narrow interval of $z$-integration ($\simeq\,10^{-10}$). The integral in Eq.~(\ref{corrPME}) also varies rapidly within this narrow interval because of the following reason. The integrand contains the Jacobi polynomial $P^{\alpha,\beta}_{K}(z)$ and its weight function $W_{\ell}(z)=(1-z)^{\alpha}(1+z)^{\beta}$ ~\cite{Abram}. For large $N$, both $P^{\alpha,\beta}_{K}(z)$ and $W_{\ell}(z)$ change very rapidly with respect to $z$. $W_{\ell}(z)$ varies from zero at $z=-1$ to a maximum of $\simeq 2^{\alpha}$ at $z_{m}=(\beta -\alpha)/(\beta +\alpha)$ and then rapidly reaching a value about $10^{-10}$ of the peak value at $z = -1 + 0.003$. Although the peak value $2^{\alpha}$ is extremely large for large $N$, partial cancellation results from the factor $[h_{K}^{\alpha \beta} \,h_{K^{\prime}}^{\alpha \beta}]^{-1/2}$~~\cite{Abram}. Thus, any standard quadrature to evaluate the integrand in Eq.~(\ref{corrPME}) gives essentially zero for $N>50$. Usually we solve this problem by splitting the interval $z\in\left[ -1, 1 \right]$ into $n$ gradually increasing subintervals and evaluating the integral in each subinterval using a 32-point Gauss-Legendre quadrature. This permits us to evaluate $V_{KK^{\prime}}(r)$ for $N$ up to 15000 with an accuracy of one part in $10^{9}$~\cite{TKD-PRA-2007}.

\subsection{Extension to $N\rightarrow \infty$}
As pointed out earlier, the experimental BEC treats up to $10^{8}$ atoms in the trap. But the numerical code mentioned above can treat only up to 15000 atoms which is far from the experimental situation. To circumvent the problem and extend the correlated many-body technique to quite a large number of atoms, we recently made a direct mathematical transformation~\cite{Sofianos}, which transforms the PHEM into a two-variable integro-differential equation. With this transformation, the Jacobi polynomial $P_{K}^{\alpha \beta} (z)$ is replaced with the associated Laguerre polynomial. In our initial attempt we applied the CPHEM using the Laguerre polynomial (CPHEL) for the order of $10^{6}$ atoms for the ground state~\cite{Sofianos}. We utilize a mathematical relation $\beta \rightarrow \infty$ to transform the Jacobi polynomials into the associated Laguerre polynomials~\cite{Abram}. An outline of the derivation, including derivation of the relations between the Jacobi and the associated Laguerre polynomials in the limit $\alpha \rightarrow \infty$ are as follows. Starting from the mathematical relation~\cite{Abram} 
\begin{equation}
  \lim\limits_{\beta \rightarrow \infty}P_n^{\alpha \beta}
   \left(1-\frac{2x}{\beta}\right)
  =L_n^{\alpha}(x),
\end{equation}
interchanging $\beta$ and $\alpha$, and using the relation~\cite{Abram}
\begin{equation}
  P_n^{\alpha \beta}(-x)=(-1)^n\,P_n^{\beta \alpha}(x)\,,
\end{equation}
we obtain
\begin{equation}
  \lim\limits_{\alpha \rightarrow \infty}P_n^{\alpha \beta}\left(\frac{2x}{\alpha}-1\right)  =(-1)^n\,L_n^{\beta}(x)\,.
\end{equation}
Substituting $x=\zeta^2=\alpha\,(r_{ij}/r)^2$ and $z=2(r_{ij}/r)^2-1$, we get
\begin{equation}
  \lim\limits_{\alpha \rightarrow \infty}P_K^{\alpha \beta}(z)=(-1)^K\,L_K^{\beta}(\zeta^2) \,.\label{Limitrel}
\end{equation}
This relation was used in evaluating $f_{K\ell}$ appearing in the CDEs for large $\alpha$~\cite{Sofianos}. In this limit, the weight function $W_{\ell}(z)$ of the Jacobi polynomial transforms as
\begin{equation}
  W_{\ell}(z) = (1-z)^{\alpha}(1+z)^{\beta} 
  = \frac{2^{\alpha+\beta}}{\alpha^\beta}\,\zeta^{2\beta}\left(1-\frac{\zeta^2}{\alpha}\right)^\alpha\,.
\end{equation}
In the limit $\alpha \rightarrow \infty$, the last factor becomes ${\rm e}^{-\zeta^2}$. Hence for large $\alpha$,
\begin{equation}
 W_{\ell}(z)=\frac{2^{\alpha+\beta}}{\alpha^\beta}\,\zeta^{2\beta}\,{\rm e}^{-\zeta^2}.  \label{Limwtfun}
\end{equation}
This has the correct functional form for the weight function of the associated Laguerre polynomial $L_K^\beta(\zeta^2)$. Substituting equations~(\ref{Limitrel}) and (\ref{Limwtfun}) in equation~(\ref{corrPME}), and using the explicit expression of the norm of the Jacobi polynomial~\cite{Abram}, we obtain 
\begin{eqnarray} 
V_{K,K^{\prime}}(r) = & A_c \int_{x_{min}}^{\alpha}L_K^{\beta}(x)\,V\left(r\sqrt{\frac{x}{\alpha}}\right)\,\eta\left(r\sqrt{\frac{x}{\alpha}}\right) 
\,L_{K^{\prime}}^{\beta}(x)\,x^{\beta}\,{\rm e}^{-x}\,{\rm d}x
\label{LaguPME}
\end{eqnarray}
where $x_{min}=(r_c/r)^2 \,\alpha$, $r_c$ is the hard-core radius of our chosen realistic van der Waals potential and 
\begin{eqnarray} 
A_c & = & \frac{(-1)^{K+K^{\prime}}}{\alpha^{\beta}}\Bigg[ 
\frac{2K+\gamma}{\alpha} \cdot\frac{2K^{\prime}+\gamma}{\alpha}\cdot \frac{\Gamma(K+1)}{\Gamma(K+\beta+1)}  \nonumber\\ 
&& \times\; \frac{\Gamma(K^{\prime}+1)}{\Gamma(K^{\prime}+\beta+1)}
 \cdot\frac{\Gamma(K+\gamma)}{\Gamma(K+\alpha+1)}\cdot \frac{\Gamma(K^{\prime}+\gamma)}{\Gamma(K^{\prime}+\alpha+1)}\Bigg]^{1/2}
\end{eqnarray} 
where $\gamma = \alpha+\beta+1$.

\section{The ground-state energy for $N \simeq 10^{7}$ atoms in an external trap}
\label{sec:egs-largeN}

Throughout our calculation we keep the system parameters fixed at values which correspond to the JILA trap~\cite{Anderson}. The mass $m=87 $ amu, the trap frequency $\omega_{ho} = 2 \,\pi \times 77.78$ Hz and the scattering length $a_s=100$ Bohr. However, for comparison with the DMC results with repulsive a hard sphere interaction, we also compute the ground-state energy for few bosons with a larger scattering lengths $a_{s} = 1000$ Bohr and $a_{s} = 10000$ Bohr. As a unit of length we choose the oscillator unit (o.u.) of length $a_{ho} =\sqrt{\hbar/(m \,\omega_{ho})}$ and the energy unit as the harmonic oscillator energy $\hbar \,\omega_{ho}$. For the mean-field GP equation, the two-body potential is chosen as the zero-range potential $V(r)= 4 \,\pi\,\hbar^2\, a_s\,\delta(r)/m$ \cite{Dalfovo}. This potential is shape-independent and completely ignores the dependence of energy on the scattering amplitude. We choose the realistic van der Waals interaction
\begin{eqnarray}
 V(r_{ij}) = \left\{\begin{array}{cl} -\frac{C_6}{r_{ij}^6} & 
\;\mbox{for}\quad r_{ij} > r_c \\
 \infty & \;\mbox{for}\quad r_{ij} \le r_c \\
\end{array}\right.          
\end{eqnarray}
with $C_6= 6.489755 \times 10^{-11}$ o.u. for $^{87}$Rb atoms~\cite{Pethick}. For a given value of $r_c$, $a_s$ is calculated by looking at the zero-energy solution of the two-body Schr\"odinger equation Eq.~(\ref{eq.tbse}), where $V(r_{ij})$ is the van der Waals potential~\cite{Pethick}. Its asymptotic form quickly attains $\eta(r_{ij}) \sim C(1 - a_s/r_{ij})$ from which $a_s$ is determined~\cite{Pethick}. We choose the value of $r_c=1.121054 \times 10^{-3}$ o.u. which corresponds to $a_s=100 \;{\rm Bohr} =0.00433\; {\rm o.u.}$. It is to be noted that although $r_c$ is almost four times smaller than $a_s$, they are of same order. Smaller values of $r_c$ (corresponding to a larger number of nodes in $\eta(r_{ij})$) are not chosen to avoid the presence of the many-body bound states and clustering. It should also be noted that for hard-sphere scattering, the effective range $r_e$ increases linearly with $a_s$, as $r_e=\frac{2}{3}a_s$, whereas for the van der Waals potential $r_e$ is determined from $a_s$ by \cite{Gao}
\begin{equation}
 \frac{r_{e}}{\beta_{6}} = \left( \frac{2}{3x_{e}}\right)\, \frac{1}
{(a_{s}/\beta_{6})^{2}}\left\{1+\left[1-x_{e}\left(\frac{a_{s}}
{\beta_{6}}\right)\right]^{2}\right\},  
\label{eq.vdw-r_e}
\end{equation}
where $\beta_{6}=(m\,C_{6}/\hbar^{2})^{1/4}$ and $x_{e} = \big[\Gamma(\frac{1}{4})\big ]^{2}/2\pi$. The calculated value of $r_{e}$ for our present work is $1.251 \times 10^{-3}$ o.u. which is comparable with but greater than the value of $r_{c}$, as expected.

\begin{figure}[hbpt]
\vspace{-10pt}
\centerline{
\hspace{-3.3mm}
\includegraphics[width=0.8\linewidth]{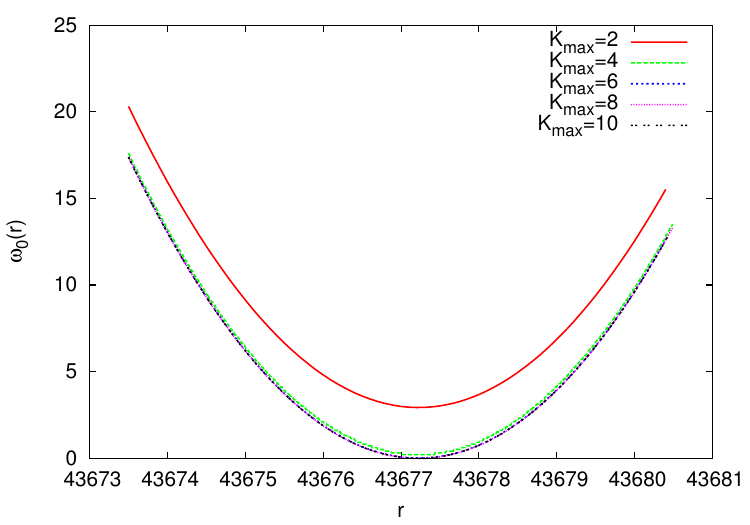}}
\caption{(color online) Plot of the effective potential $\omega_0$ (in o.u.) as a function of the hyperradius $r$ (in o.u.) for different 
values of $K_{max}$ for $10^7$ atoms in the trap. We have set the zero of the ordinate at the minimum $\omega_{min}$ ($ = 794854262.3813$ o.u.)
of the effective potential $\omega_0(r)$ for $K_{max}=10$.}
\label{fig.pot}
\end{figure}

As pointed earlier, the lowest eigen potential $\omega_0(r)$ is treated as the many-body effective potential which describe the collective phenomena of a dilute BEC. This is also in good agreement with experimental situation. As at the ultra-cold temperature all the individual atoms in the condensate lie within a single de-Broglie wavelength, the condensate is treated as a single lump of quantum stuff. However, before calculating the ground-state energies, it is indeed required to check the convergence for $N = 10^{7}$ atoms. With the increase in the number of particles $N$, the effective interaction $N\,a_s$ increases, the condensate becomes more repulsive. The condensate density is pushed out as its average radius increases sharply. So higher $K_{max}$ may be needed for convergence of the ground state for larger $N$. In Fig.~\ref{fig.pot} we plot $\omega_0(r)$ as a function of $r$ for $K_{max}=2,4,6,8$ and $10$ with $N=10^7$ and observe a very fast convergence. Although after $K_{max}=4$, all the graphs appear to overlap completely as seen in the Fig.~\ref{fig.pot}, still we noticed that the minimum of the effective potential decreases very slowly (which is not visible in the figure) as $K_{max}$ increases. This is consistent with the Rayleigh-Ritz principle. So, throughout our calculation we fix $K_{max}=8$ and calculate the ground state energy per particle. In Table~\ref{table:egs}, we compare our many-body results with the TF, GP and MGP results for different diluteness of the condensate given by $n\,a_s^3$. We evaluate the gas parameter $n\,a_{s}^{3}$ at the center of the trap which can be directly expressed in terms of relevant parameters as \cite{Dalfovo}
\begin{equation}
n(0)\,a_{s}^{3} = \frac{15^{2/5}}{8\,\pi} 
\left(N^{1/6} \,\frac{a_{s}}{a_{ho}} \right)^{12/5}
\end{equation}
Note that the value of $n(0)\,a_{s}^{3}$ is deliberately kept $<<1$ to justify the usage of the two-body correlated basis functions. Comparison with the MGP is needed for better justification as the MGP includes the correction due to quantum fluctuations. 
\begin{table}[h!]
\caption{Ground-state energy per particle (in o.u.) for $^{87}$Rb atoms. Results in the TF approximation, solving the GP, MGP and the CPHEL (CPHEM using the Laguerre polynomial) are presented in unit of ($\hbar\omega_{ho}$). The results of the correlated Hartree HNC equation [Eq.~(17) of Ref~\cite{Poll}] are also included.} \label{table:egs}
\begin{tabular}{ccccccc}
\hline\noalign{\smallskip}
 $N$ &  $n(0)\,a_{s}^{3} $ & TF  & GP  & MGP & CPHEL & HNC \\[3pt]
\noalign{\smallskip}\hline\noalign{\smallskip}
 $10^3$&   $3.961 \times 10^{-6}$  & 1.90   & 2.42   & 2.43   & 2.43   & 2.43 \\
 $10^4$&   $9.949 \times 10^{-6}$  & 4.76   & 5.04   & 5.08   & 5.19   & 5.04 \\
 $10^5$&   $2.499 \times 10^{-5}$  & 11.96  & 12.10  & 12.25  & 12.67  & 12.20 \\
 $10^6$&   $6.277 \times 10^{-5}$ & 30.05  & 30.12  & 30.66  & 31.67  & 30.48 \\
 $10^7$&   $1.577 \times 10^{-4}$ & 75.49  & 75.52  & 77.48  & 79.48  & 76.85 \\
\noalign{\smallskip}\hline
\end{tabular}
\end{table}

We calculate the GP energy by solving the standard GP energy functional 
\begin{eqnarray}
E_{GP}[\Psi]= & \int {\rm d}\vec{r} \,\Big[\frac{\hbar^2}{2\,m}\,\big|\,\vec{\nabla}\,\Psi\,\big|^{2}+\frac{1}{2}\,m\omega_{ho}^2 \,r^2\, \big|\,\Psi\,\big|^{2} + \frac{2\,\pi\,{\hbar}^{2}\,a_s}{m}\,\big|\,\Psi\,\big|^{4}\Big]\,. \label{eq.gp-functional}
\end{eqnarray} 
and the MGP energy is calculated by solving 
\begin{eqnarray}
E_{MGP}[\Psi] = & \int {\rm d}\vec{r} \Bigg[\frac{\hbar^2}{2\,m}\,\big|\,\vec{\nabla}\, \Psi\,\big|^{2} + \frac{1}{2}\,m\,{\omega_{ho}}^{2}\,{r}^{2}\,\big|\,\Psi\,\big|^{2} + \nonumber\\
&\qquad \frac{2\,\pi\,{\hbar}^{2}\,a_s}{m}\,\big|\,\Psi\,\big|^{4}\,\Big(1+\frac{128\, {a}_{s}^{3/2}}{15\sqrt{\pi}}\,\big|\,\Psi\,\big|\Big)\Bigg] \,. \label{eq.mgp-functional}
 \end{eqnarray}
The additional term in Eq.~(\ref{eq.mgp-functional}) basically adds quantum corrections to the mean-field effective potential. The TF energy is calculated by using a simple analytic expression~\cite{Dalfovo}. All the results are presented in Table~\ref{table:egs}. 

We observe that our many-body ground state energy is almost indistinguishable from those of both the GP and MGP for small $N$, as expected. However, for larger $N$, we observe that our many-body results are consistently closer to the MGP results. Note that the TF results are always lower than both the GP and many-body results as the kinetic energy term is completely ignored in the TF limit. We calculate the correlation energy \cite{PRA2014} 
$$
E_{corr}^{GP} = \frac{E_{\rm {many{\mbox -}body}}-E_{\rm GP}}{E_{\rm {many{\mbox -}body}}}
$$ 
as a measure of the small deviation of the mean-field GP and the TF results from our many-body results. 
\begin{figure}[hbpt]
\vspace{-10pt}
\centerline{
\hspace{-3.3mm}
\includegraphics[width=0.8\linewidth]{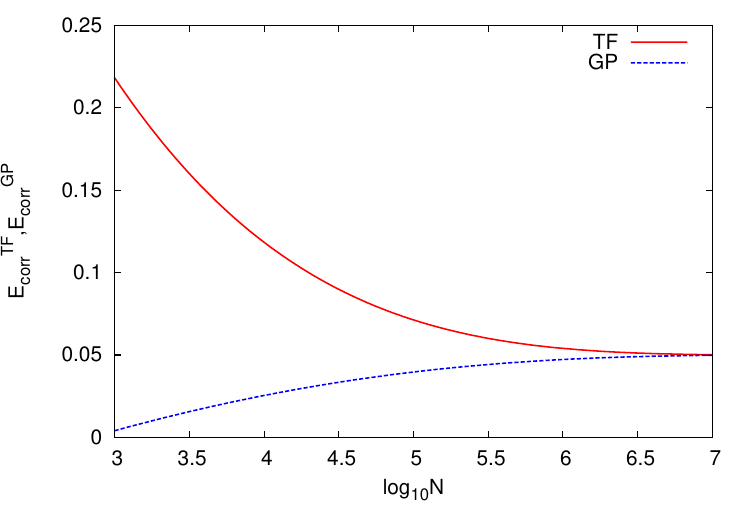}}
\caption{(color online) The normalized correlation energy calculated using the TF ($E_{corr}^{TF}$) and the GP ($E_{corr}^{GP}$) energies as a function of $\log N$.}\label{fig.corr-energy}
\end{figure}
We present our results in Fig.~\ref{fig.corr-energy} and observe that both $E_{corr}^{TF}$ and $E_{corr}^{GP}$ converge to the same very small value at the very large particle-number limit. This reaffirms that the ground state properties of a dilute BEC in the truly thermodynamic limit should be correctly described by the TF equation. In the same table (Table~\ref{table:egs}), we also compare our results with the correlated Hartree HNC results (solution of eq. (17) of Ref~\cite{Poll}) which utilizes hard sphere bosons. We find that our many-body energies are larger than the HNC results. The presence of a hard-core part in the van der Waals interaction produces an excluded volume. Therefore, for the same $N$, our many-body method should produce larger clusters than the GP and the HNC, and hence, has a larger contribution from the harmonic trapping potential. Of course there is also the effect of the attractive tail of the van der Waals potential which tries to reduce the energy. However, in the dilute condition it does not have a significant contribution which can be detected experimentally. Therefore, the combined effect of a hard core and an attractive tail along with the presence of a centrifugal repulsion term in the coupled differential equations make the CPHEL ground state energy more repulsive. As expected, this term will have a larger contribution for larger $N$ as shown in Table~\ref{table:egs}.

Before addressing the validity of the SIA for large $N$, it is instructive to have an estimation of the accuracy of our two-body correlated basis function method. Thus we compare the CPHEM results with the available DMC results~\cite{Blume}, which are essentially exact. For completeness we also include the GP and the MGP results. We have presented the total ground state energies for various numbers of bosons $N$ interacting via a repulsive hard-sphere potential both for weak and strong interaction in Table~\ref{table:dmc}. For $a_{s} = 100$ Bohr, the CPHEM results are in good agreement with the DMC results which guarantees the applicability of two-body correlated basis function in the dilute regime. It further assures us to carry forward the calculations for large $N$ where the DMC approach fails and our correlated basis functions approach works. It is also noted from Table~\ref{table:dmc} that the relative difference between the DMC and the CPHEM results is almost of the order of $10^{-3}$ for $a_{s} = 100$ Bohr. Although we are not interested in the strong interaction in the present work, we also report results for larger $a_s$, for completeness. The relative difference between the DMC and the CPHEM results is again of the order of $10^{-3}$ for $a_{s} = 1000$ Bohr while for $a_{s}$ = 10000 Bohr the relative difference is of the order of $10^{-1}$ or more. We may conclude that for stronger interactions, the inclusion of higher body correlations may be required. 
\begin{table}[h!]
\caption{The ground-state energy (in o.u.) for $^{87}$Rb atoms. The $s$-wave scattering length $a_s$ is also in o.u.. Results of the DMC ~\cite{Blume}, the GP, the MGP and the CPHEM are also presented. The DMC and the CPHEM results are reported for the repulsive hard-sphere interaction.}
\label{table:dmc}
\begin{tabular}{ccccccc}
\hline\noalign{\smallskip}
 $a_s$ & $N$ &  DMC &  GP  & MGP & CPHEM  \\
\noalign{\smallskip}\hline\noalign{\smallskip}
100 & 3  &  4.510  &  4.510 & 4.510 & 4.506 \\ 
    & 5  &  7.534  &  7.534 & 7.534 & 7.531 \\ 
    & 10 & 15.153  &  15.153 & 15.153 & 15.143 \\ 
    & 20 & 30.640  &  30.638 & 30.639 & 30.622 \\ \noalign{\smallskip}
1000 & 3  & 4.603 & 4.600 & 4.602 & 4.575 \\ 
     & 5  &7.835 & 7.826 & 7.834 & 7.775\\ 
     & 10 & 16.426 & 16.383 & 16.426 & 16.280 \\ 
     & 20 & 35.475 & 35.293 & 35.497 & 35.180 \\ \noalign{\smallskip}
10000 & 3 & 5.553  & 5.329 & 5.611 & 5.055 \\ 
      & 5 & 10.577  & 9.901 & 10.722 & 9.471\\ 
      & 10  & 26.22 & 23.61 & 26.84 & 23.072\\ 
      & 20  & 66.9 & 57.9 & 68.5 & 50.678\\
\noalign{\smallskip}\hline
\end{tabular}
\end{table}

\section{The shape-independent approximation}
\label{sec:SIA}

To conclusively address the issue of validity of the SIA, a detailed study of the ground-state energies over a wide range of the $C_{6}$ parameter is required. Although some papers~\cite{TKD-PRA-2008,Esry1,Hau,Esry2,Blume} present interesting discussions of this issue, none of them considers the wide range of $N$ and none use the realistic inter-atomic interaction. Considering few tens of atoms in the trap, it is shown that the many-body results approach closer to the mean-field results when the number of particles $N$ in the trap increases. It can be intuitively understood that tthe system becomes more classical with increasing $N$. In fact, the SIA is valid for a truly finite number of atoms (few tens) and in the extremely dilute condition. However, our present calculation starts with only a few atoms and goes up to the order of millions of atoms in the trap. This requires a deeper and thorough study of the validity of the SIA for such a wide range of $N$. Using the van der Waals interaction instead of the hard-sphere interaction, it is very easy to study the effect of the long-range attractive tail in the ground-state properties of the BEC and to observe its universal behavior.

We tune the $C_{6}$ parameter and by changing the cutoff radius $r_{c}$, we fix the scattering length $a_{s}$ to 100 Bohr, which mimics the JILA experiment. The calculated many-body results together with the GP and MGP results are presented in Table~\ref{table:SIA}. Again we observe that the many-body ground state energy is almost indistinguishable for up to a few thousands of atoms. For larger $N$, the ground-state energies deviate slightly and the relative change is almost negligible. To have an estimation, we consider the relative change for two extreme choices of $C_{6}$. When the value of $C_{6}$ for these two choices was varied by a factor of $\frac{9}{5}=1.8$, the ground-state energy for the largest choice of the number of atoms ($N= 10^{7}$) varied by a factor of $\frac{79.8}{79.2}$ = $1.0073$. This indicates that the ground-state energy is almost insensitive to the exact value of $C_{6}$ for a wide range of $N$ within the given density parameter $n\,a_{s}^{3} << 1$. Thus, the ground-state energy considerably satisfies the SIA.   

\begin{table}[!h]
\caption{%{\label{tableX}
Ground-state energies per particle calculated by CPHEL (CPHEM using the Laguerre polynomial) in o.u. for different values of $C_{6}$ for $3\leq N \leq 10^{7}$ atoms in the Rb condensate. All values of $C_{6}$ (given in o.u. in the top line of columns 2 to 6) correspond to the same scattering length $a_{s} = 100$ Bohr. The GP and MGP results are presented for comparison.} \label{table:SIA}
\begin{tabular}{llllllll}
\hline\noalign{\smallskip}
$N$ & $ 5 \times 10^{-11}$ &  $6.489 \times 10^{-11}$ & $7\times 10^{-11}$  &   $8\times 10^{-11}$ & $9\times 10^{-11}$ & GP & MGP\\ 
\noalign{\smallskip}\hline\noalign{\smallskip}
3  &  1.512  & 1.512   & 1.512  &  1.512   & 1.512 & 1.511 & 1.511\\
5  &  1.519  & 1.519   & 1.519  & 1.519    & 1.519 & 1.514  & 1.515\\
10 & 1.522   & 1.522   & 1.522  & 1.522    & 1.522 & 1.517 & 1.519\\
20 & 1.548   & 1.547   & 1.547  & 1.547    & 1.547 & 1.533 & 1.535\\
100 & 1.678  & 1.677   & 1.677  & 1.676    & 1.676  & 1.652 & 1.653\\
$10^{3}$  & 2.435  & 2.434 & 2.432 & 2.432 & 2.430 & 2.424 &  2.43\\
$ 10^{4}$ & 5.199  & 5.198 & 5.194 & 5.188 & 5.182 & 5.08 & 5.08\\
$10^{5}$  & 12.70  & 12.68 & 12.66 & 12.64 & 12.62 & 12.10 & 12.25\\
$10^{6}$  & 31.75  & 31.68 & 31.64 & 31.60 & 31.56 & 30.12 & 30.66\\
$10^{7}$  & 79.80   & 79.56 & 79.42 & 79.31 & 79.22 & 75.52 & 77.48\\
\noalign{\smallskip}\hline
\end{tabular}
\end{table}

\begin{figure}[hbpt]
\vspace{-10pt}
\centerline{
\hspace{-3.3mm}
\includegraphics[width=0.8\linewidth]{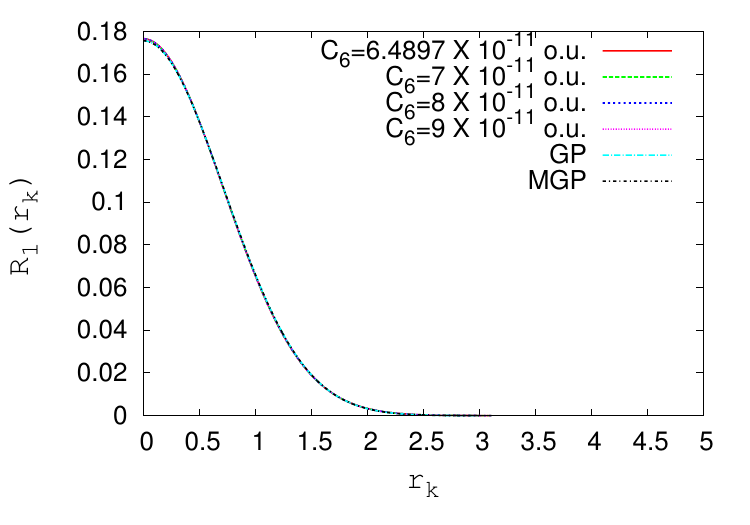}}
\caption{(color online) Plot of the one-body density $R_1(\vec{r}_k)$
for $N=10$ $^{87}$Rb atoms in the condensate for various $C_6$ values corresponding to the same $a_s=100$ Bohr ($n\,a_s^3 \sim 10^{-6}$). The corresponding GP and MGP results are also presented for comparison. $R_1(\vec{r}_k)$ is calculated using Eq.~(\ref{eq.onebd-def}) with $\Psi$ obtained from the CPHEL, GP and MGP respectively. }
\label{fig.onebd}
\end{figure}

\begin{figure}[hbpt]
    \begin{tabular}{cc}
    \includegraphics[width=0.45\linewidth]{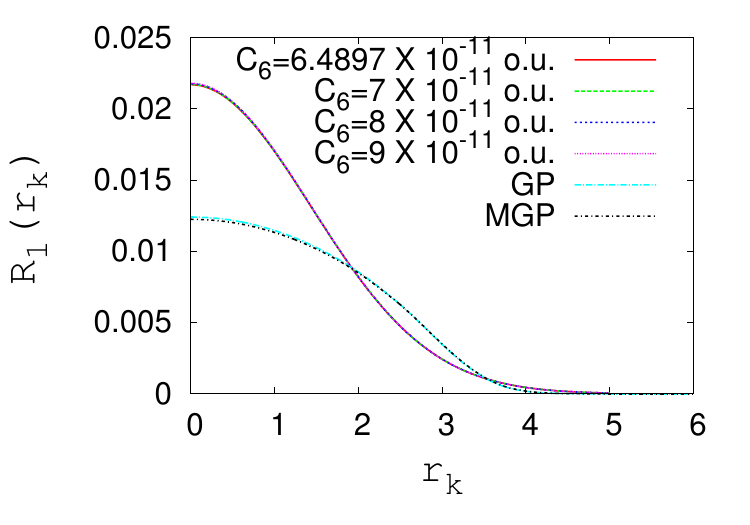}&
    \includegraphics[width=0.45\linewidth]{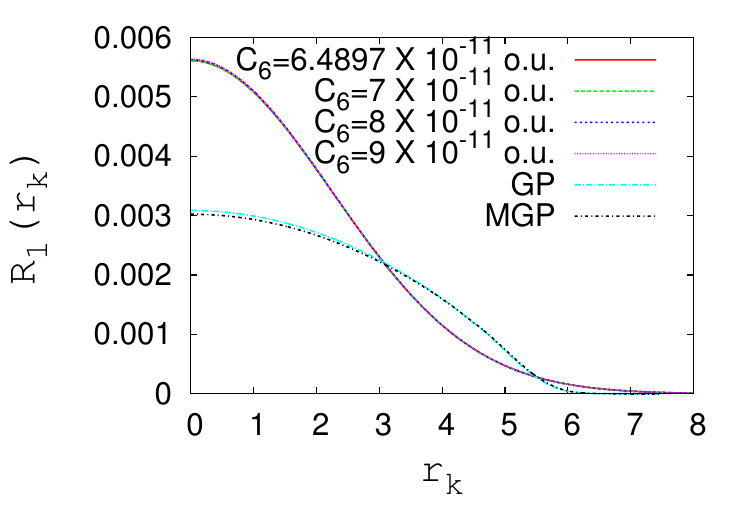}\\
             (a)  &        (b)   \\
    \includegraphics[width=0.45\linewidth]{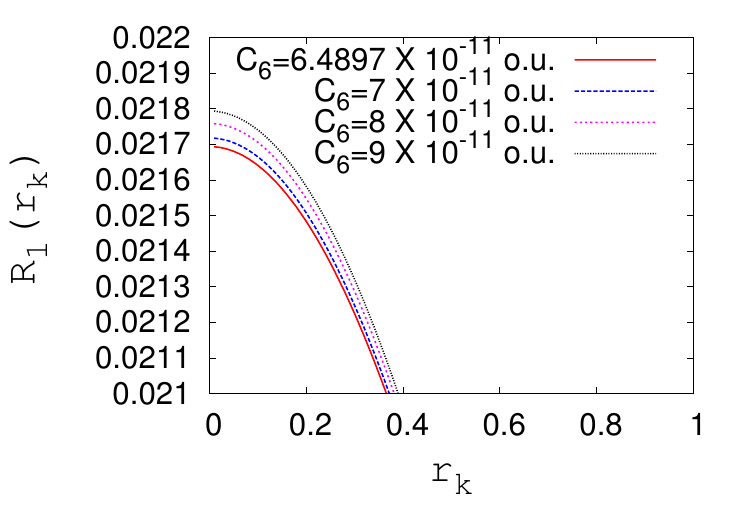}&
    \includegraphics[width=0.45\linewidth]{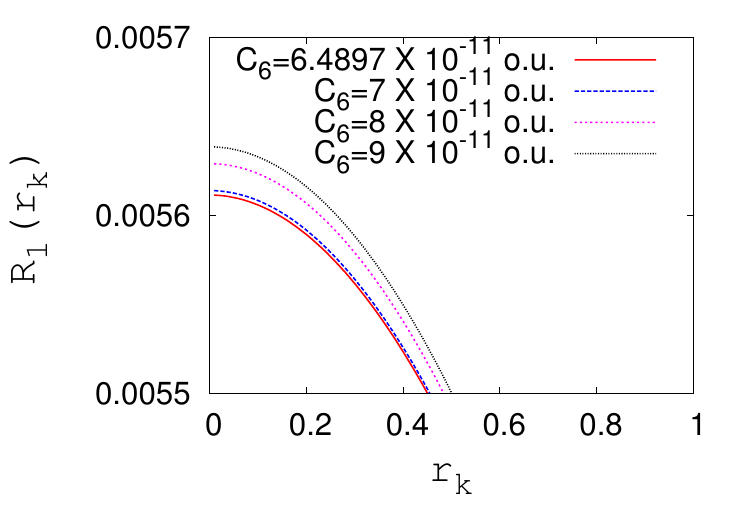}\\
         (c)  &     (d)\\  
    \includegraphics[width=0.45\linewidth]{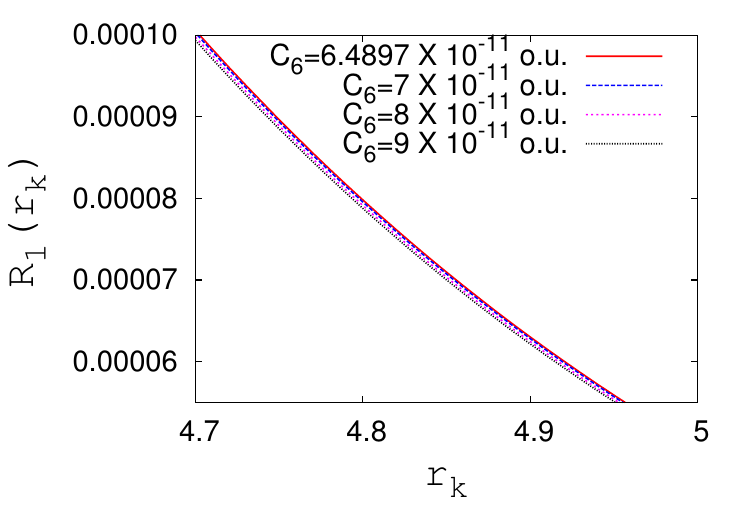}&
    \includegraphics[width=0.45\linewidth]{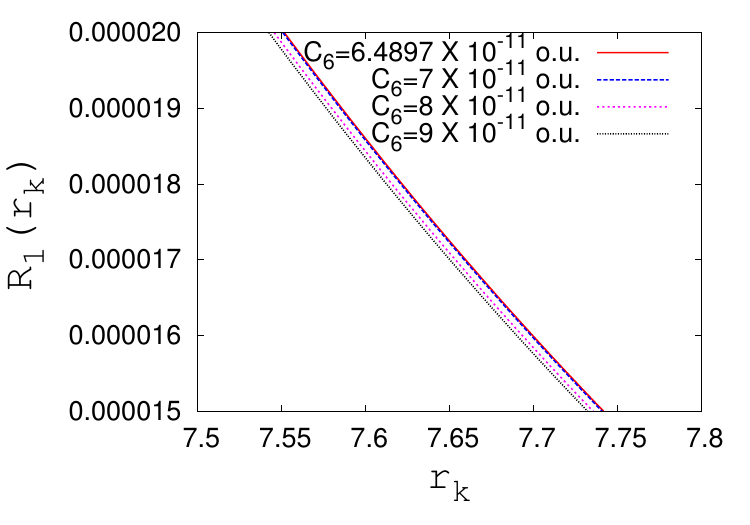}\\
              (e) &      (f)\\
    \end{tabular}
\caption{(color online) One-body densities $R_1(\vec{r}_k)$ for various $C_6$ corresponding to the same $a_s=100$ Bohr are plotted for (a) $N=10000$ ($n\,a_s^3 \sim 10^{-3}$) and (b) $N=100000$ ($n\,a_s^3 \sim 10^{-4}$). For comparison the corresponding GP and MGP results are also plotted. $R_1(\vec{r}_k)$ is calculated using Eq.~(\ref{eq.onebd-def}) with $\Psi$ obtained from the CPHEL, GP and MGP respectively. To highlight the effect of shape dependence of the interacting potential we present the enlarged view of the peak portions of the curves in panel (c) (for $N = 10000$) and (d) (for $N =100000$), and the corresponding tail portions in the panel (e) and panel (f) respectively.}
\label{fig.SIA}
\end{figure}

Next we compare the one-body density for various $C_{6}$ parameters and for a wide range of $N$. Although we have reported some results on the one-body density for smaller $N$ and have observed an appreciable effect of finite size, in the present work we are interested in the thermodynamic limit. The one-body density is a key quantity as it contains information regarding one-particle aspect of the condensate and can be indirectly measured in the interferometry experiments. We define it as the probability density of finding a particle at a distance $r_k$ from the centre of mass of the condensate~\cite{JCP2010}
\begin{equation}
R_1(\vec{r}_k)=\int_{\tau^{\prime}}|\,\Psi\,|^2 {\rm d}\tau^{\prime}
\label{eq.onebd-def}
\end{equation}
where $\Psi$ is the full many-body wave function and the integral over the hypervolume $\tau^{\prime}$ excludes the variable $\vec{r}_k$. In Fig.~\ref{fig.onebd} we present the calculated one-body density for $N=10$ atoms in the trap and compare with the GP results. The calculated one-body density for various values of $C_{6}$ perfectly match with the GP results which confirms our previous observations. In Fig.~\ref{fig.SIA} we plot the one-body density for $N = 10^{4}$ atoms (panel (a)) and for $N=10^{5}$ atoms (panel (b)) in the trap, and compared with the GP results. The density profiles calculated from the GP equation have the same qualitative features, however, disagreement remains in the peak value of the density distribution as well as in the extension of the density profile, which will be discussed later. We plot the enlarged profile of the one-body density near the peak [panel(c) for $ N=10^{4}$ and panel(d) for $N=10^{5}$] and near the tail part [panel(e) for $N=10^{4}$ and panel(f) for $N=10^{5}$] for various $C_{6}$ parameters. It is seen that all the many-body results calculated for various $C_{6}$ and having the same scattering length $a_{s}$ are almost indistinguishable and may not be detected in an experiment. This means that in the dilute regime $(n\,a_{s}^{3} << 1)$, the condensate is well described by the single parameter $a_{s}$. 

\begin{figure}[hbpt]
\vspace{-10pt}
\centerline{
\hspace{-3.3mm}
\includegraphics[width=0.8\linewidth]{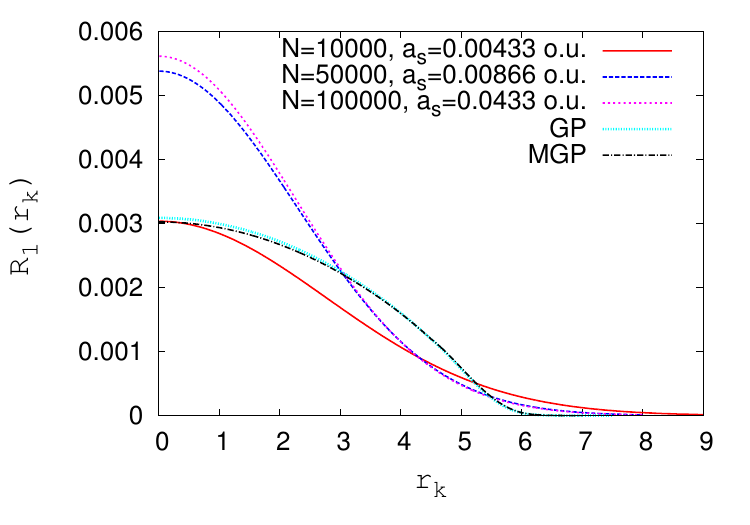}}
\caption{(color online) Plot of the one-body density $R_1(\vec{r}_k)$ for a fixed $N\,a_s$ but different combinations of $N$ and $a_s$. The corresponding GP and MGP results are also plotted for comparison. $R_1(\vec{r}_k)$ is calculated using Eq.~(\ref{eq.onebd-def}) with $\Psi$ obtained from CPHEL, GP and MGP respectively. }\label{fig.onebd-na}
\end{figure}

The disagreement between the many-body results and the GP results at the peak value needs additional discussion for which we plot the one-body density in Fig.~\ref{fig.onebd-na}, having the same value of the effective repulsive interaction $N\,a_{s}$, but different choices of $N$ and $a_{s}$. The actual two-body attraction is determined by the integration $4\pi \int_{r_c}^{\infty} \,V(r)\, \eta(r)\, r^2\, {\rm d} r$. In a many-body calculation which uses the van der Waals potential having a long attractive tail $-C_{6}/r^{6}$, the net effective interaction is more attractive than in the GP case. Being more repulsive, the GP treatment lowers the central density and expands the density distribution. It is clearly seen that for the choice of $a_{s} = 0.0433\; {\rm o.u.} = 1000\; {\rm Bohr}$ and $N$ = 10000, the many-body results perfectly agree with the GP and MGP results at the peak value. Thus keeping $N\,a_{s}$ constant, one can make the effect of repulsion stronger by appropriately choosing $N$ and $a_{s}$. By gradually increasing $a_{s}$, the effect of the repulsive interaction increases in our many-body calculation, the central peak gradually shifts downward and extended outward, while the mean-field results are independent of the separate choices of the number of atoms and scattering length. This supports the observed effect of using finite number of atoms in the many-body calculations, which has been discussed earlier in many contexts.

\section{Collective excitations at low energy}
\label{sec:collective}

\begin{figure}
    \begin{tabular}{cc}
    \includegraphics[width=0.45\linewidth]{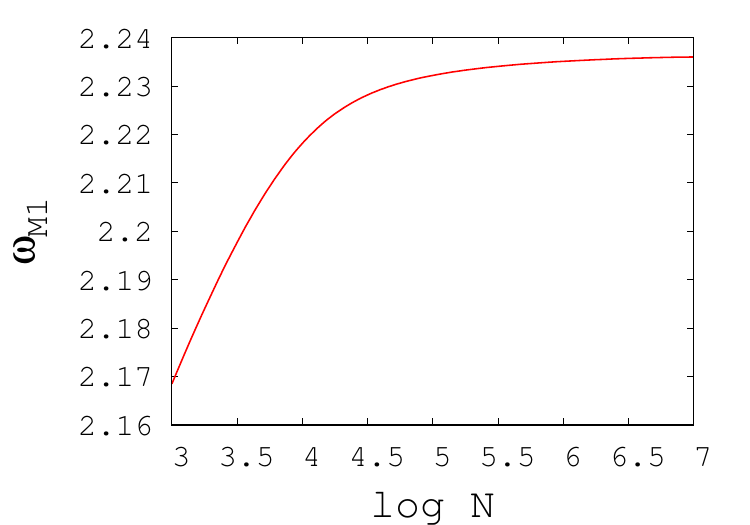}&
    \includegraphics[width=0.45\linewidth]{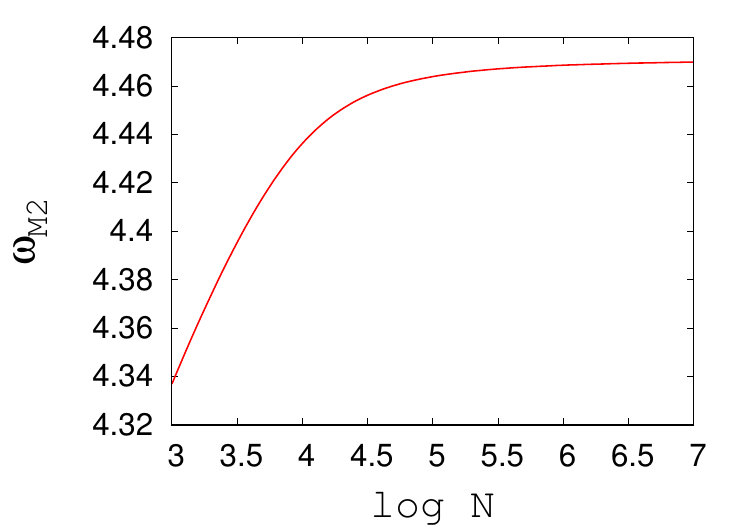}\\
     (a) monopole & (b) second breathing mode \\
    \includegraphics[width=0.45\linewidth]{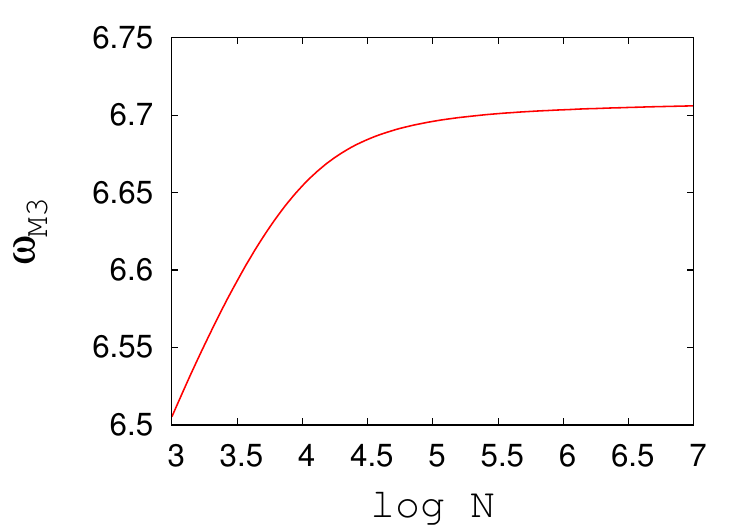}&
    \includegraphics[width=0.45\linewidth]{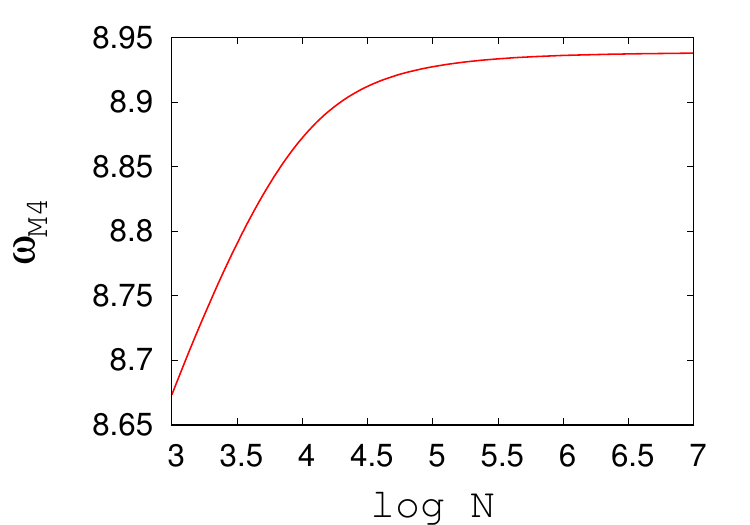}\\
     (c) third breathing mode & (d) fourth breathing mode \\
    \end{tabular}
\caption{(color online) Plot of different breathing mode frequencies as a function of $\log N$.} \label{fig.breathing-mode}
\end{figure}

It was already pointed out that the low-energy collective excitations provide valuable information about the interaction, while the high-lying excitations are of single particle nature and are useful for the study of statistical properties. In our present picture, the collective motion of the condensate in the hyperradial space takes place in the effective potential $\omega_0(r)$. The ground state in this well gives the ground-state energy $E_{00}$ of the condensate corresponding to $n=0$ and $\ell=0$. We use the notation $E_{n\ell}$ for the energy (in o.u.) of the $n$-th radial excitation of the $\ell$-th surface mode. The monopole frequency $\omega_{M1}$ is defined as the lowest hyperradial excitation corresponding to the breathing mode ($\ell=0$) and is calculated as $\omega_{M1} =E_{10}-E_{00}$. For $\ell \neq 0$ we get the surface modes which can be calculated as hyperradial excitations in the eigen potential $\omega_\ell(r)$ corresponding to different values of $\ell$. However, large inaccuracy relating to the calculation of the off-diagonal potential matrix elements for $\ell \neq 0$ and difficulties with the slow converge present challenges. For large $N$, we observed that the diagonal hypercentrifugal term is very large which contributes most to the potential matrix. Hence we disregard $\ell > 0$ contributions to the off-diagonal matrix elements and construct the effective potential $\omega_\ell(r)$ in the hyperradial space for non-zero orbital angular momentum. In Fig.~\ref{fig.breathing-mode} we plot several breathing mode excitation frequencies like $\omega_{M1}=E_{10}-E_{00}$ (monopole), $\omega_{M2}=E_{20}-E_{00}$ (second breathing mode), $\omega_{M3}=E_{30}-E_{00}$ (third breathing mode) and $\omega_{M4}=E_{40}-E_{00}$ (fourth breathing mode) as functions of $\log N$. In the description of the breathing mode frequencies at the large $N$ limit, it is important to calculate and compare frequencies of the different modes using the hydrodynamic (HD) model~\cite{Stringari,Dalfovo}
\begin{equation}
\omega(n,\ell)=\omega_{ho}(2\,n^2+2\,n\,\ell+3\,n+\ell)^{1/2} .
\label{eq.HD}
\end{equation}
where $\ell$ and $n$ are the angular momentum quantum number and number of nodes in the radial solution, respectively. 
For quite a large $N$, when the dimensionless parameter $N\,a_s/a_{ho}$ is large, the kinetic energy term in the ground state GP equation becomes negligibly small compared to mean-field term and one gets the TF approximation~\cite{Dalfovo}. In the same limit the eigen frequencies are calculated using the hydrodynamic (HD) equation of superfluidity (Eq.~(\ref{eq.HD})). So the comparison of the many-body results to the HD prediction is well justified. Note that the HD equation (Eq.~(\ref{eq.HD})) depends only on $n$ and $\ell$ and not explicitly on $N$. Thus the effect of finite size correction does not appear here and the many-body results should coincide with HD results in the true thermodynamic limit although small deviation may exist for finite size system. In Fig.~\ref{fig.breathing-mode}, all the breathing mode frequencies saturate at the large $N$ limit. The asymptotic values for the several breathing modes are presented in Table~\ref{table:breathing-mode} and compared with the HD results.
\begin{table}[h]
\caption{Comparison between the asymptotic values of several breathing mode frequencies calculated by the CPHEL and the HD predictions.}
\label{table:breathing-mode}
 \begin{tabular}{lcccc}
\hline\noalign{\smallskip}
 $\omega_{Mn}$ & CPHEL & HD \\ 
\noalign{\smallskip}\hline\noalign{\smallskip}
 $\omega_{M1}$  & 2.236 & 2.236 \\ %$\sqrt{5}$
 $\omega_{M2}$  & 4.468 & 3.742 \\ %$\sqrt{14}$
 $\omega_{M3}$  & 6.702 & 5.196 \\ %$\sqrt{27}$
 $\omega_{M4}$  & 8.936 & 6.633 \\ %$\sqrt{44}$
\noalign{\smallskip}\hline
\end{tabular}
\end{table}
It is seen that the HD prediction is very accurate for the description of the lowest excitation of large systems. However, we observe gradually increasing deviations as we go to higher modes. The slow but smooth increase in $\omega_{Mn}$ with increase in $N$ is visible in all four panels of Fig.~\ref{fig.breathing-mode}, which basically manifests the finite size effect. The deviation from the HD results for larger $N$ can be attributed to the following reason. In the calculation of $\omega_{l}(r)$ for $l>0$ we have taken the contributions from the diagonal matrix element and neglected the off-diagonal matrix elements as we face numerical difficulty as discussed earlier. Though their contribution is practically insignificant compared to diagonal part, however, this approximation may not be true for higher modes.

\begin{figure}[hbpt]
\vspace{-10pt}
\centerline{
\hspace{-3.3mm}
\includegraphics[width=0.8\linewidth]{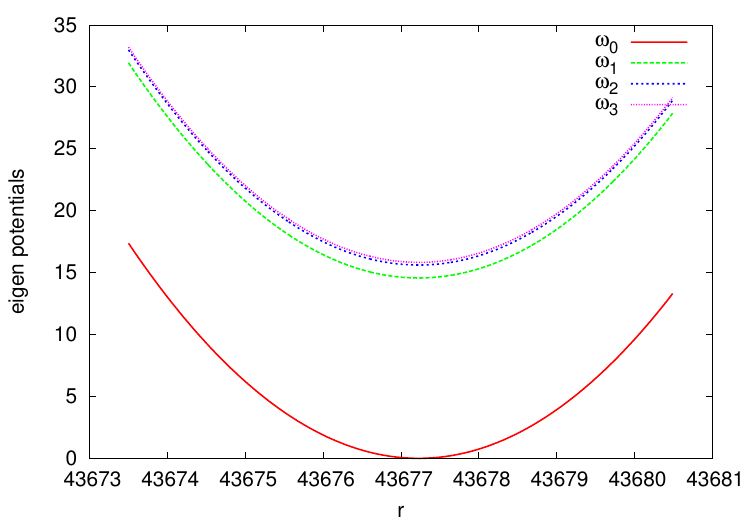}}
\caption{(color online) Plot of the higher order (lowest four) eigen potentials $\omega_\ell$ (in o.u.) for $N=10^{7}$. We have set the zero of the ordinate at the minimum $\omega_{min}$ ($ = 794854262.3821$ o.u.) of the lowest eigen potential $\omega_0(r)$.} \label{fig.eig-pot}
\end{figure}

Another aspect of the many-body calculation is to calculate the high-lying excitations which are of single particle nature. However, as mentioned earlier, we calculate the excitation spectrum by using the eigen potential and for the present calculation we consider $\omega_{0}$ as the lowest eigen potential. This approximation is quite justified for the ground state and the low-lying excitations as shown in Fig.~\ref{fig.eig-pot}. For high-lying excitations, the effect of higher eigen potential may come in the picture and for the accurate calculation of the high-lying excitations we must use the coupled adiabatic channels. Therefore, for the present manuscript we leave the calculation of high-lying excitations to future work.

\section{Summary and conclusion}
\label{sec:conclusion}

We have calculated the ground-state energy, the one-body density and the low-lying excitation frequencies for a very large number of trapped bosons, which is close to the real experimental situation. We utilized the two-body correlated basis functions and used the van der Waals interaction as the inter-atomic interaction. The many-body method described in this manuscript can reveal realistic features of the trapped bosons. The most convenient and widely used tool in this direction is the mean-field GP equation which basically ignores inter-atomic correlation and uses the simple contact interaction. In that respect our many-body approach is a few steps ahead of the mean-field approach as by keeping all possible two-body correlations one can expect to address the beyond the mean-field effects. On the other hand the diffusion Monte-Carlo method is the exact many-body technique. However, due to computational difficulties it can handle only up to few hundreds of bosons in the trap, which is far from the real experimental situation. Our many-body method keeps only two-body correlations and can handle as many as $10^7$ atoms in the trap. The effect of only two-body correlations is relevant as the higher-body correlations are almost negligible in a dilute BEC.

In the first part of our calculation we applied the many-body approach for the calculation of ground-state energies. Comparison with the mean-field results through the correlation energies with respect to the GP and TF results demonstrate that the BEC looses its many-body effects and becomes more classical at a truly large-particle limit. This can be understood from the fact that for a large effective repulsion (large $\frac{Na_s}{a_{ho}}$ with $a_s>0$), particles are far apart from each other and the effect of interaction becomes small. The TF approximation well describes such a situation. Our present calculation also deals with the wide range of particle numbers and present an exhaustive study of the validity of the shape-independent approximation. However, another fundamental motivation of the present work is to study the collective excitations. We calculate several excited modes of breathing mode frequencies and compare with the HD model. We observe that the asymptotic value ($N \rightarrow \infty$) of the lowest breathing mode (monopole frequency) exactly matches with the HD prediction. Whereas the higher breathing modes ($\omega_{Mn}, n=2,3,4$) have the same qualitative nature as $\omega_{M1}$, the asymptotic values are a bit higher than that of HD prediction. For smaller $N$, the finite size effects exist and the breathing modes show slow and smooth increase until it approaches the asymptotic value ($N \rightarrow \infty$). We conclude that the low-lying collective excitations are well described by the HD model at $N \rightarrow \infty$. Our present calculation is an exhaustive study of both the static and dynamic behavior of trapped bosons. However, in the present work we strictly confine our attention to diluteness of the order of $n\,a_{s}^{3} <<1.0$ as we keep the effect of two-body correlations. Thus we leave the study of higher density regime to future works.

\begin{acknowledgement}
BC acknowledges the financial support of Department of Science and Technology (DST), Govt. of India, under a Major Research Project [Sanc. No. SR/S2/CMP-126/2012]. 
\end{acknowledgement}

\end{document}